\documentclass[]{JHEP3}
\usepackage{epsfig}

\def\be{\begin{equation}}
\def\ee{\end{equation}}
\def\beq{\begin{equation}}
\def\eeq{\end{equation}}
\def\beqar{\begin{eqnarray}}
\def\eeqar{\end{eqnarray}}
\def\barr{\begin{array}}
\def\earr{\end{array}}
\def\lsim{\:\raisebox{-0.5ex}{$\stackrel{\textstyle<}{\sim}$}\:}

\def\and{\qquad {\rm and } \qquad}

\def\slp{p \hspace{-1ex}/}
\def\slk{$k$ \hspace{-1ex}/}

\def\slk{k \hspace{-1ex}/}

\def\ie{{\it i.e.}}

\def\sbar{ \overline{s} }
\def\thmin{\theta_0}
\def\cmin{\cos \theta_0}

\title{\boldmath New physics in $e^+e^- \to Z \gamma$ with polarized beams}
\author{Balasubramanian Ananthanarayan\\Centre for High Energy Physics,
Indian Institute of Science, Bangalore 560 012, India \\ 
E-mail: 
\email{anant@cts.iisc.ernet.in}}
\author{Saurabh D. Rindani\\ Theory Group, Physical Research Laboratory,
Navrangpura, Ahmedabad 380 009, India \\
E-mail: \email{saurabh@prl.ernet.in}}
\abstract{We present a complete description of angular distributions 
in the presence of new interactions for the process
$e^+ e^-\to Z \gamma$ with polarized beams
at future linear colliders, by considering the most general form-factors
allowed by gauge invariance.  We include the possibility of CP
violation, and classify the couplings according to their 
CP properties. Chirality conserving and chirality 
violating couplings give rise to distinct dependence on beam
polarization.  We present a comprehensive discussion including 
both types of couplings and provide a detailed comparison of the effects
due to each. We discuss some selected asymmetries 
which would enable isolating effects of the CP-violating form-factors.
We also present sensitivities on the corresponding couplings
achievable at a future linear collider with realistic polarization and
luminosity.  }

\preprint{PRL-TH-05-3}

\keywords{Beyond standard model, CP violation} 

\begin{document}

\section{Introduction}\label{intro}
An international linear collider (ILC) that will
collide $e^+$ and $e^-$ at centre of mass energy of $\sqrt{s}=
500~{\rm GeV}$ is now a distinct possibility.  The aim of this
machine would be to determine the parameters of the standard
model (SM) at higher precision than ever before, discover the
Higgs boson and establish its properties, produce particles
that have so far not been accessible at present day energies,
and probe physics even due to interactions mediated by particles
that are too massive to be produced.  
One important window for
the observation of beyond the standard model (BSM) physics is to establish
CP violation outside the neutral meson system.  As a result, it
is important to discuss all the physics possibilities that would
lead to such CP violation, the imprint each type of
interaction would leave on measurements in as model independent
a manner as possible, and to advance new and improved tools
to probe such interactions.  Thus, any work, like ours, 
which discusses the impact of new interactions should 
lay special  emphasis on CP-violating observables. 

It is by now clear that the availability of polarization of the beams, 
both longitudinal and transverse, would play an indisputable role in enhancing 
the sensitivity of observables to CP violation, and indeed would play 
complementary roles \cite{Moortgat-Pick:2005cw}. For instance, the availability of longitudinal 
polarization would enhance the possibility of detecting possible 
dipole moments of, say the $\tau$-lepton~\cite{ars} and the 
top-quark~\cite{cr}.  For the important process of $e^+ e^-\to
t\overline{t}$, the availability of transverse polarization would allow one to 
probe BSM scalar and tensor type 
interactions~\cite{ar1}.  
The last result was explicitly demonstrated some time ago, 
and could have been, in principle, deduced from the distinguished work of 
Dass and Ross~\cite{dr}.  The considerations here arise from
properties of the interference of standard 
model amplitudes, which at tree-level
are generated by the exchange of $Z$ and $\gamma$ in the $s-$channel with
amplitudes represented by contact interactions due to 
BSM interactions.  We note here that for light quarks,
a discussion was presented in earlier work~\cite{br}.

On the other hand, $Z\gamma$
production, a process for which there is a significant SM cross-section
(for early work, see e.g. \cite{Renard:1981es}),
presents surprises since the SM contribution arises from the $t-$ and
$u-$channel contributions, rather than $s-$channel contributions 
as in the top-quark case.  In a recent article~\cite{ar2}, we discussed
at length the contributions to the differential cross-section due
to completely model independent, most general gauge and Lorentz invariant,
chirality conserving (CC) contact interactions.  
[Contact interactions have also been considered in the context of
$s-$channel processes in, e.g., ref.~\cite{Grzadkowski:1997cj,BOPP,BDDR}.]
In particular, one of these
contact interactions generates precisely the same contributions as
those generated by anomalous CP violating triple-gauge boson vertices
studied in a similar context somewhat earlier~\cite{Choudhury:1994nt,arsb},
when the parameters are suitably identified.  
In this paper, we complement the work of \cite{ar2} by including a
discussion of chirality violating (CV) contact interactions for new
physics. The discussion here is thus comprehensive and includes the sum
total of all Lorentz and gauge invariant interactions beyond the
standard model. It thus
provides a platform for a model independent discovery of BSM physics.

The term contact interactions, as we use it here, calls for some
explanation. The term is usually used in a low-energy effective 
theory with a cut-off
energy scale $\Lambda$ for effective interactions induced in the form of
nonrenormalizable terms by new physics at some high scale, and this
consists of an expansion in inverse powers of $\Lambda$. The expansion
is then terminated at some suitable inverse power of $\Lambda$, keeping
effective higher-dimensional operators arising from new interactions
up to a certain maximum dimension. In our approach we do not introduce a
cut-off, nor do we limit the dimensions of the operators. We simply write
down all independent forms of amplitudes in momentum space relevant for
the process in question, with coefficients which are Lorentz invariant
form-factors, and are functions of the kinematic invariants. Thus, in
principle, we keep all powers of momenta, unlike in the low-energy
effective theory approach. 

Our formalism thus encompasses
not only the standard contact interactions, but, for example, also
interactions where there may be propagators for the exchange of new or
SM virtual particles in the $s$, $t$ or $u$ channel. 

We thus use the
term contact interactions for convenience, to denote a general
form-factor approach, in the spirit of Abraham and Lampe~\cite{AL},
in momentum space for the amplitude for the process
in question (for a recent mention see~\cite{PR-Z}).
It is worthwhile emphasizing that in the present framework,
$\Lambda$ does not play the role of a cut-off;
therefore, the scale $\Lambda$ will make no explicit appearance in our 
considerations.  As a result, we will scale all BSM dimensionful
parameters by the only mass scale in the problem which is $m_Z$, while
noting that the latter only plays the role of a book-keeping device. 
To avoid confusion, we should state 
that even though we do not introduce explicitly a large
energy scale $\Lambda$, our form-factors need not necessarily be of
order unity -- they may be suppressed by couplings constants and/or inverse
powers of some heavy masses.

In the absence of general
results {\it \`a la} Dass and Ross for interactions of this type, it is not
possible to guess what would happen if there were model independent
form factor representation for BSM physics due to
CV  interactions.
These could, in principle, be generated by either 
scalar type interactions containing no Dirac $\gamma$ matrices or a
$\gamma_5$ matrix, or tensor type interactions  
containing an anti-symmetrized product of two $\gamma$ matrices,
$\sigma_{\mu \nu}$.  It turns out that in the 
limit of vanishing
electron mass, the contributions of the latter always reduce to certain
combinations indistinguishable from those induced by scalar type interactions.
While our explicit computations realize
this, which we do not report here, it may be seen to follow from some
general considerations which we prove.  
Therefore, the linearly independent set of form-factors we employ here
are those that are of the scalar type.  

After presenting the results from the scalar form factors, 
we construct some sample asymmetries and evaluate them
in terms of the new interactions. This enables us to provide examples of 
limits
on the sensitivities that can be achieved in future collider experiments.
We then discuss models in which such interactions can be generated.
In addition to presenting our novel results, here we also present
a comprehensive discussion on all the issues involved, including
recounting important aspects of established results.

We begin with a general discussion on chirality
conservation and violation in general and the role of transverse
polarization in uncovering interactions of each type in Sec.~\ref{general}.
This discussion parallels the one presented by Hikasa~\cite{Hikasa},
which we shall briefly outline.
We then specialize to the process of interest in Sec.~\ref{zgamma}.
We present a detailed discussion and summarize our conclusions
in Sec.~\ref{discconc}. We provide proof of the redundancy
of tensor form-factors in Appendix~\ref{appa}, while
Appendix \ref{appb} contains a discussion of the CP properties of
various couplings in the contact interactions.

\section{Chirality conservation and violation}\label{general}

Polarization effects are different for new interactions which are 
chirality conserving and for those which are chirality violating. 
Firstly, in the limit of vanishing electron mass, 
there is no interference of the chirality violating new interactions 
with the the SM interactions, since the latter are chirality conserving. As a 
result, there is no contribution from chirality violating interactions 
which is polarization independent or dependent on longitudinal 
polarization in the limit of vanishing electron mass. 
Transverse polarization effects for the two cases are also different. 
The cross terms of the SM amplitude with the amplitude from chirality 
conserving contact interactions has a part independent of transverse 
polarization and a part which is bilinear in transverse polarization of 
the electron and positron, denoted by $P_T$ and
$\overline P_{T}$ respectively. For the case of chirality violating 
interactions, the cross term has only terms linear in $P_{T}$ and 
$\overline P_{T}$, and no contributions independent of these.

The features discussed above
have been captured in the work of Hikasa~\cite{Hikasa},
wherein the cross-section $\Sigma$ for the transversely polarized
case is expressed as 
\begin{eqnarray*}
&\displaystyle \Sigma=\Sigma_{\rm unpol}-
{1\over 2} P_T {\overline P_{T}} {\rm Re}\, [T^*_{++} T_{--}]+
{1\over 2} P_T \overline P_{T}  {\rm Re}\, [e^{-2i\phi_H} T^*_{+-}T_{-+}]+
					& \\
& \displaystyle {1\over 2} P_T {\rm Re} [e^{-i\phi_H}\left(T^*_{+-} T_{--}+
				       T^*_{++} T_{-+} \right)] -
{1\over 2} {\overline P_T} {\rm Re} [e^{-i\phi_H}(T^*_{++} T_{-+} 
						   +T^*_{--} T_{-+})], 
						& 
\end{eqnarray*}
where the $T_{++},\, T_{+-},\, T_{-+}$ and $T_{--}$ are helicity
amplitudes for the process at hand, $\phi_H$ is the final-state azimuthal
angle and $\Sigma_{\rm unpol}$ is the unpolarized cross-section. 
In the $T_{ab}$, $a,b=+,-$, the subscripts stand for the helicities
of the $e^+$ and $e^-$ respectively.  In the above, 
BSM interactions of the CC type contribute to
the amplitudes $T_{+-}$ and $T_{-+}$, while those of the CV 
type contribute to $T_{aa},a=+,-$ (the SM interactions themselves
contribute only to $T_{+-}$ and $T_{-+}$, when $m_e$ effects are neglected).  
Note also the characteristic $2\phi_H$ dependence accompanying the
terms bilinear in transverse polarization, and the $\phi_H$ dependence
accompanying the linear transverse polarization pieces when the
BSM physics is worked out to leading order.
In this manner, the general treatment 
of Hikasa provides a useful reference guide to the polarization
dependence of the cross-sections on BSM interactions, and can be
used to check the observations of the preceding paragraph. 

Due to the above-mentioned dependence on transverse polarization, 
the type of CP violating observables is also different in the two cases of 
chirality conservation and chirality violation.
It is thus clear that the two cases should be treated separately, and
this is what we do in the following.  From hereon, we shall specialize
to the process of interest in this work, {\it viz.} $Z\gamma$
production.

\section{\boldmath The process $e^+ e^- \to Z \gamma$ with contact 
interactions}\label{zgamma}
In this section we consider the  process 
\begin{equation}
e^-(p_-,s_-)+e^+(p_+,s_+)\rightarrow \gamma (k_1,\alpha)+Z(k_2,\beta),
        \label{process}
\end{equation}
parametrize the contribution to its amplitude in terms of form-factors
introduced in a contact-interaction description of physics beyond the
standard model, and discuss asymmetries in the consequent angular
distributions.

\subsection{Contact interactions with chirality conservation and violation}

We shall assume that the amplitudes are generated by the standard model
as well as a general set of form-factors of the type
proposed by Abraham and Lampe \cite{AL} (see also
ref.~\cite{AL2}).  They are completely 
determined by vertex factors that we denote by $\Gamma^{SM}_{\alpha\beta}$,
$\Gamma_{\alpha\beta}^{CC}$   
and $\Gamma_{\alpha\beta}^{CV}$ in a self-explanatory notation.
Of these, $\Gamma_{\alpha\beta}^{CV}$ is being proposed here for the first time.

The vertex factor  corresponding to SM is given by
\begin{equation}
\Gamma^{SM}_{\alpha\beta}={e^2\over 4 \sin \theta_W \cos \theta_W}
 \left\{\gamma_\beta 
(g_V-g_A\gamma_5)
\frac{1}{\slp_- - \slk_1} 
\gamma_\alpha+
\gamma_\alpha 
\frac{1}{\slp_- -\slk_2}                
\gamma_\beta 
(g_V-g_A \gamma_5) 
\right\}.
\end{equation} 
In the above, the vector and axial vector $Z$ couplings of the electron are 
\begin{equation}
 g_V= -1 + 4\sin^2\theta_W ;\quad g_A = -1
     \label{gVgA}.
\end{equation}
The chirality conserving anomalous form factors may be introduced 
via the following
vertex factor, which is denoted here by:
\begin{eqnarray}\label{anomcc}
& \displaystyle \Gamma_{\alpha\beta}^{CC}={i e^2 \over 4 \sin\theta_W \cos\theta_W}
\left\{\frac{1}{m_Z^4}\left((v_1+  a_1 \gamma_5)\gamma_\beta(
2 p{_-}_\alpha (p_+\cdot k_1)-
2 p{_+}_\alpha (p_-\cdot k_1)) + \right. \right. & \nonumber \\ 
& \displaystyle \left. \left.
 ((v_2+  a_2 \gamma_5) p{_-}_\beta + 
(v_3+  a_3 \gamma_5) p{_+}_\beta)(\gamma_\alpha 2 p_-\cdot k_1-
2 p{_-}_\alpha \slk_1)+
\right. \right.& \nonumber \\
& \displaystyle \left. \left.
 ((v_4+  a_4 \gamma_5) p{_-}_\beta +
(v_5+  a_5 \gamma_5) p{_+}_\beta)(\gamma_\alpha 2 p_+\cdot k_1-
2 p{_+}_\alpha \slk_1)\right)+ \right. & \nonumber \\
& \displaystyle \left. \frac{1}{m_Z^2} 
(v_6+  a_6 \gamma_5)(\gamma_\alpha k_{1\beta}-\slk_1 
g_{\alpha\beta}) 
 \right\} .&
\end{eqnarray}
This was discussed by us earlier in \cite{ar2}.

We now introduce the corresponding CV form-factors.  
In terms of a linearly independent set of scalar
form-factors (i.e., ones with no Dirac $\gamma$'s), the vertex factor can
be written down as:
\begin{eqnarray}\label{anomcv}
& \displaystyle 
\Gamma_{\alpha\beta}^{CV}={e^2\over 4 \sin\theta_W \cos\theta_W} \cdot & 
\nonumber \\
& \displaystyle \left(
(s_1  + i p_1 \, \gamma_5) [(k_1\cdot k_2)g_{\alpha\beta}-k_{1\beta} \, 
k_{2\alpha}]/m_Z^3 + \right. & \nonumber \\
& \displaystyle \left.
(s_2  + i p_+ \, \gamma_5)p_{-\beta}((k_1\cdot p_+) p_{-\alpha} - 
(k_1\cdot p_-) p_{+\alpha})/m_Z^5+
\right. & \nonumber \\
& \displaystyle \left.
(s_3  + i p_3 \, \gamma_5)p_{+\beta}((k_1\cdot p_+) p_{-\alpha} - 
(k_1\cdot p_-) p_{+\alpha})/m_Z^5+
(s_4  + i p_4 \, \gamma_5)\epsilon_{\alpha\beta\rho\sigma} k_1^{\rho} k_2^{\sigma}/m_Z^3+
\right. & \nonumber \\
& \displaystyle \left.
(s_5  + i p_5 \, \gamma_5)\epsilon_{\alpha\beta\rho\sigma} k_1^{\rho}(p_-- p_+)^{\sigma}/m_Z^3+
(s_6  + i p_6 \, \gamma_5)[\left(2(k_1\cdot p_-)(k_1\cdot p_+)\epsilon_{\alpha\beta\sigma\tau}
\right .\right. & \nonumber \\
& \displaystyle \left. \left.
\left.+\epsilon_{\beta\rho\sigma\tau} k_1^\rho (p_{-\alpha}(k_1\cdot p_+)
+p_{+\alpha} (k_1\cdot p_-))\right) p_-^\sigma p_+^\tau]/m_Z^7 \right. \right) &
\end{eqnarray}

\TABLE{
\begin{tabular}{||c|c||}\hline
 CP even & CP odd\\ \hline
$v_1$ &  \\
$v_2-v_5$ & $v_2+v_5$ \\
$v_3-v_4$ & $v_3+v_5$ \\
& $v_6$ \\
$a_1$ &  \\
$a_2-a_5$ & $a_2+a_5$ \\
$a_3-a_4$ & $a_3+a_5$ \\
& $a_6$ \\ \hline
\end{tabular}
\caption{CP even and odd combinations
of CC couplings}
}
The form-factors introduced above are in principle functions of the
Lorentz invariant quantities $s$ and $t$. In a specific frame (as for
example the $e^+e^-$ centre-of-mass frame which we employ) they could be
written as functions of $s$ and $\cos\theta$, where $\theta$ is the
production angle of $\gamma$. However, in what follows, we will assume
for simplicity 
that the form-factors are all constants. To the extent that we restrict
ourselves to a definite $e^+e^-$ centre-of-mass energy, the absence of
$s$ dependence is not a strong assumption. However, the absence of
$\theta$ dependence is a strong assumption, and relaxing this assumption
can have important consequences, as discussed below.
We begin by recalling that
one can write form-factors, which are functions of the
Mandelstam variables
$s, t, u$ as functions of just $s$ and $t-u$, since
\begin{equation}
s+t+u = m_Z^2.
\end{equation}
Further recalling that
$t-u = (s-m_Z^2) \cos\theta,$ our form-factors are, in general,
functions $F_i (s,\cos\theta)$ of $s$ and $\cos\theta.$
Furthermore, we can write each form-factor as a sum of
even and odd parts as
\begin{equation}
F_i (s, \cos\theta) = f_i (s, \cos\theta) + g_i (s, \cos\theta),
\end{equation}
where the $f_i$ are even and $g_i$ odd functions of $\cos\theta$.
Note that $\cos\theta$ changes sign under CP.
Hence, if $F_i$ occurs with a certain tensor which is CP conserving
(violating), then the $f_i$ part contributes an amount which is CP
conserving (violating), and the $g_i$ part an amount which is CP violating
(conserving). In principle, our analysis can be done taking the $f_i$ 
and $g_i$ in to
account. However, it would be extremely complicated given the large
number of form factors. 

\TABLE{
\begin{tabular}{||c|c||}\hline
 CP even & CP odd\\ \hline
$s_1$ &  \\
$s_2-s_3$ & $s_2+s_3$ \\
& $s_4$ \\
$s_5$ & \\
$s_6$ & \\
& $p_1$ \\
$p_2+p_3$ & $p_2-p_3$ \\
$p_4$& \\
& $p_5$ \\
& $p_6$ \\ \hline
\end{tabular}
\caption{CP even and odd combinations of CV couplings} }

As noted in \cite{ar2}, the combinations 
$v_1, v_2-v_5, v_3-v_4$ and $a_1, a_2-a_5, a_3-a_4$
are CP conserving, while $v_2+v_5, v_3+v_4, v_6$ and 
$a_2+a_5, a_3+a_4, a_6$ and are CP
violating, assuming that they are functions only of $s$.
As for the form-factors for the CV interactions, the combinations $s_1,
s_2-s_3, s_5, s_6, p_2+p_3$ and $p_4$ are CP conserving and $s_2+s_3,
s_4, p_1, p_2 - p_3, p_5$ and $p_6$ are CP violating, again assuming
them to be functions of $s$ alone.
(In Tables 1 and 2 we present the CP properties of
all the combinations of couplings of interest.)
In Appendix \ref{appb} we demonstrate how the CP properties of the
couplings may be determined and derive the consequences of the CPT
theorem.  

A discussion is in order on the number of CP violating form-factors we expect
to have.  Given that the $Z$ is a massive vector particle and that the photon
is a massless one, and that the electron is a spin 1/2
particle, we have 12 helicity amplitudes.  We can, therefore, 
have 12 
form-factors in the chirality conserving case, neglecting the electron
mass.  Of the $v_i,\, a_i, i=1,...,6$, only three
linear combinations of each are CP violating, and since each is complex, the
total number is 12 as expected, and so also for the CP conserving case.
The count is analogous also for the chirality violating case considered here
with $s_i,\, p_i, i=1,...,6$.

\subsection{The differential cross-section}
We now give expressions for the differential cross-sections including
both CC and CV contact interactions for polarized beams.
The differential cross-section for longitudinal beam polarizations $P_L$ and 
$\overline P_L$ of  $e^-$ and $e^+$  is given by
\be
\displaystyle
\left(\frac{d\sigma}{d\Omega    }\right)_L =
    {\cal B}_L\left(1-P_L\overline{P}_L\right)
\left[
       \frac{1}{\sin^2 \theta}
          \left( 1 + \cos^2 \theta + \frac{4 \sbar}{( \sbar - 1)^2}
           \right)
     + C_L              
\right]  \; ,
    \label{diff c.s.L}
\ee
where
\begin{eqnarray}
& \displaystyle
\sbar  \equiv  \frac{s}{m_Z^2},\,\,
   {\cal B}_L  = \frac{\alpha^2}{16 \sin^2\theta_W m_W^2 \sbar}
     \left( 1 - \frac{1}{\sbar}   \right)
     (g_V^2+g_A^2-2Pg_Vg_A) , & 
\end{eqnarray}
where the effective polarization parameter $P$ is defined as
\be
P = \frac{P_L - \overline{P}_L}{1- P_L \overline{P}_L},
\ee
and
\begin{eqnarray}
& \displaystyle
C_{L}  =  
        \frac{1}{4 (g_V^2+g_A^2-2Pg_Vg_A)}
    \left\{\sum_{i=1}^6
\left(	(g_V-Pg_A) {\rm Im}v_i+ (g_A-Pg_V) {\rm Im}a_i\right) X_i 
\right\},&
\end{eqnarray}
and where the $X_i,i=1,...,6$ are kinematical
quantities resulting from our computations and are listed in Table 3
(along with $Y_i,i=1,...,6$ and $Z_i,i=1,...,6$ which enter
transversely polarized cross-sections in the following). 
Note that our convention is one in which positive $P_{L}$ corresponds to 
right polarized electrons and positive $\overline P_{L}$ to right polarized 
positrons. 
The differential cross-section for transverse polarizations $P_T$ and 
$\overline{P}_T$ of $e^-$ and $e^+$ is given by
\begin{eqnarray}
& 
\displaystyle
\left(\frac{d\sigma}{d\Omega}\right)_T =
    {\cal B}_T
\left[
       \frac{1}{\sin^2 \theta}
          \left( 1 + \cos^2 \theta + \frac{4 \sbar}{( \sbar - 1)^2}
	- P_T \overline{P}_T\frac{g_V^2-g_A^2}{g_V^2+g_A^2}
	\sin^2 \theta \cos 2\phi 
           \right)+ \right. & \nonumber \\
& \displaystyle \left.
      C_T^{CC}+C_T^{CV}              
\right]  \; ,
    \label{diff c.s.T}
\end{eqnarray}
with, 
\begin{eqnarray}
& \displaystyle
   {\cal B}_T  = \frac{\alpha^2}{16 \sin^2\theta_W m_W^2 \sbar}
     \left( 1 - \frac{1}{\sbar}   \right)
     (g_V^2+g_A^2) , & 
\end{eqnarray}
and
 \begin{eqnarray}
& \displaystyle
C_{T}^{CC}  =
        \frac{1}{4 (g_V^2+g_A^2)}
    \left\{\sum_{i=1}^6
        (g_V {\rm Im}v_i+ g_A {\rm Im}a_i) X_i +\right. & \nonumber \\
        & \displaystyle
    \left. P_T \overline{P}_T\,
\sum_{i=1}^6\left(
        (g_V {\rm Im}v_i- g_A {\rm Im}a_i)
    \cos 2\phi+
        (g_A {\rm Re}v_i- g_V {\rm Re}a_i)
                 \sin 2\phi\right)Y_i  \right\}, &
   \label{notation}
\end{eqnarray}
and
\begin{eqnarray}
& \displaystyle C_T^{CV}={1\over 8 (g_V^2+g_A^2)} \cdot & \nonumber \\
& \displaystyle
\left( \sum_{i=1}^{3} \left\{
(P_T-\overline P_{T}) \left[
g_V({\rm Im}\, s_i) \sin\phi - g_A ({\rm Re}\, s_i) \cos\phi \right] +
 \right. \right. & \nonumber \\
& \displaystyle \left. \left.
(P_T+\overline P_{T}) \left[
g_A({\rm Re}\, p_i) \sin\phi + g_V ({\rm Im}\, p_i) \cos\phi \right] \right\}
\cdot Z_i + \right. & \\
& \displaystyle
\left. \sum_{i=4}^{6} \left\{
(P_T-\overline P_{T}) \left[
g_V({\rm Im}\, s_i )\cos\phi + g_A ({\rm Re}\, s_i) \sin\phi \right] +
 \right. \right. & \nonumber \\
& \displaystyle \left. \left.
(P_T+\overline P_{T}) \left[
g_A({\rm Re}\, p_i) \cos\phi - g_V ({\rm Im}\, p_i) \sin\phi \right] \right\}
\cdot Z_i \right).
\end{eqnarray}
As expected,
there is no contribution from the CV interactions to the cross-section
(\ref{diff c.s.L}) with longitudinal polarization.

\TABLE{
\begin{tabular}{||c|c|c|c||}\hline
$i$ &  $X_i$ & $Y_i$ & $Z_i$\\ \hline \hline
$1$ & $-2  \sbar (\sbar+1)$ & 0 & $- 4 \sqrt{\bar s} \cot\theta$  \\  \hline
$2$ & $  \sbar (\sbar-1) (\cos\theta-1)$ & $0$ & $-\bar s^{3/2} (\bar s-1) \sin\theta/2$ \\ \hline
$3$ & $  0 $ & $\sbar(\sbar-1)(\cos\theta-1)$ & $-\bar s^{3/2} (\bar s-1) \sin\theta/2$  \\ \hline
$4$ & $  0 $ & $\sbar(\sbar-1)(\cos\theta+1)$ 
& $4 \sqrt{\bar s} \csc\theta$   \\ \hline
$5$ & $ \sbar(\sbar-1) (\cos\theta+1)$ & $0$ 
& $-4 \sqrt{\bar s} \cot\theta$ \\ \hline
$6$ & $2  (\sbar-1) \cos\theta $ & $2 (\sbar-1)\cos\theta$ 
& $-\sbar^{3/2} (\sbar-1)^2 \sin 2\theta/8$
\\ \hline
\end{tabular}
\caption{The contribution of the new couplings to the cross section}}

In the expressions above,
$\theta$ is the angle between photon and the $e^-$ directions, and $\phi$
is the azimuthal angle of the photon, with the 
$e^-$ direction chosen as the $z$
axis, and with the direction of its 
transverse polarization chosen as the $x$ axis.
The $e^+$ transverse polarization direction is chosen anti-parallel to the 
$e^-$ transverse polarization 
direction\footnote{This was incorrectly stated as 
``parallel" in \cite{ar2,arsb}.}, 
and corresponds to the convention adopted by Hikasa~\cite{Hikasa}.

We have kept only terms of leading order in the anomalous couplings, 
since they are expected to be small. The above expressions may be 
obtained either 
by using standard trace techniques for Dirac spinors with a
transverse spin four-vector, 
or by first calculating helicity amplitudes and then writing
transverse polarization states in terms of 
helicity states. 
We note that the contribution of the interference between 
the SM amplitude and the anomalous amplitude 
vanishes for $s=m_Z^2$. The reason
is that for $s=m_Z^2$ the photon in the final state 
is produced with
zero energy and momentum, and for the photon four-momentum $k_1=0$, 
the anomalous contribution  (\ref{anomcc}) vanishes identically.

\subsection{Asymmetries}
We now formulate angular asymmetries which can help to determine
different independent linear combinations of form-factors. The number of
form-factors in either the CC case or the CV case is 12. A glance at
Table 3 reveals that the number of independent angular distributions in
either case is not that large. Thus, even if electron and positron
polarizations can be turned on or off, or allowed to change signs, it
would not be possible to determine all the form-factors from an
experimental determination of the polarized angular distributions. 

We discuss below some selected asymmetries which can be useful. Our choice of
asymmetries would be ideally suited to determine the CP-violating
combinations of form-factors in the case when one could choose electron
and positron polarizations to be equal in magnitude but opposite in
sign, so that the initial state has a definite CP transformation. In
such a case, it may be checked using the CP properties of form-factors
detailed in Sec. (2.1) that the form-factors appearing in the
asymmetries chosen below are precisely in the combinations which are odd
under CP. In all cases, we use a cut-off $\theta_0$ in the forward and backward
directions on the polar angle $\theta$ of the photon. This cut-off is needed
since no observation can be made too close to the beam direction.
Moreover, it can serve a further purpose that the sensitivity can be
optimized by choosing a suitable cut-off.

For the case of the CC, we introduce the following asymmetries~\cite{ar2}:
\begin{eqnarray}& \displaystyle A_1^{CC}(\thmin)=
{1\over \sigma_0}
\sum_{n=0}^3 (-1)^n
\left(
\int_{0}^{\cos \theta_0} d \cos\theta
 -
\int_{-\cos \theta_0}^{0} d \cos\theta \right) 
 \int_{\pi n/ 2}^{\pi(n+1)/  2} d\phi \,
{d \sigma \over d \Omega} , 
\end{eqnarray}
\begin{eqnarray}& \displaystyle A_2^{CC}(\thmin)=
{1\over \sigma_0}
\sum_{n=0}^3(-1)^n \left(
\int_{0}^{\cos \theta_0} d \cos\theta -
\int_{-\cos \theta_0}^0 d \cos\theta \right)
 \int_{\pi (2 n-1)/4}^{\pi(2 n+1)/4} d\phi \,
{d \sigma \over  d \Omega} , &
\end{eqnarray}
and
\begin{eqnarray}
& \displaystyle A_3^{CC}(\thmin)=\frac{1}{\sigma_0}
\left(\int_{-\cos \theta_0}^{0} d \cos\theta-
\int^{\cos \theta_0}_0 d \cos\theta\right)
\int_{0}^{2 \pi} d\phi \,
{d \sigma \over d \Omega            }\, , &
\end{eqnarray}
with
\begin{eqnarray}
& \displaystyle \sigma_0 \equiv \sigma_0(\thmin)=
\int_{-\cos \theta_0}^{\cos \theta_0} d \cos\theta
\int_{0}^{2 \pi} d\phi \,
{d \sigma \over d \Omega            }\, . &
\end{eqnarray}
Of the asymmetries above, $A_1^{CC}$ and $A_2^{CC}$ 
exist only in the presence of transverse polarization. The asymmetry
$A_3^{CC}$, on the other hand is enhanced in the presence of
longitudinal polarization, compared to when the beams are unpolarized.
These are readily evaluated and read as follows:
\begin{eqnarray}
& \displaystyle A_1^{CC}(\theta_0)= & \nonumber \\
&  \displaystyle               
         {\cal B}'_T       
	\,  P_T \overline{P}_T\, \left[
g_A \left\{\sbar 
({\rm Re} v_3+{\rm Re}v_4)+2 {\rm Re} v_6\right\}- 
g_V \left\{\sbar 
({\rm Re} a_3+{\rm Re}a_4)+2 {\rm Re} a_6\right\}\right], 
		 \; & \\ 
 & \displaystyle A_2^{CC}(\theta_0)= & \nonumber \\
& \displaystyle               
         {\cal B}'_T  
	\,  P_T \overline{P}_T\, \left[
g_V \left\{\sbar 
({\rm Im} v_3+{\rm Im}v_4)+2 {\rm Im} v_6\right\}- 
g_A\left\{\sbar 
({\rm Im} a_3+{\rm Im}a_4)+2 {\rm Im} a_6\right\}\right], 
		 \; 
		 \end{eqnarray}
where, we have defined
\begin{equation}
{\cal B}'_T =
\frac{{\cal B}_T (\sbar - 1)\cos^2\theta_0}  
	{ (g_V^2+g_A^2)\sigma_0^T},
\end{equation}
with
\begin{eqnarray}
& \displaystyle \sigma_0^T=4 \pi {\cal B}_T
  \left[ \left\{ \frac{\sbar^2 + 1}{(\sbar - 1)^2}
                   \ln \left( \frac{1 + \cmin }
                                   {1 - \cmin }
                            \right)
                 - \cmin
          \right\}\right]. \; 
\end{eqnarray}
We note here that $A_1^{CC}$ ($A_2^{CC}$) are sensitive to real (imaginary)
parts of $v_3, \, a_3, \, v_4,\, a_4,\, v_6$ and  $v_6$.
Indeed, as mentioned in
earlier, these asymmetries bring in the transverse $e^+$
and $e^-$ polarizations bilinearly in accordance with the
general expectations presented in Sec.~\ref{general}.

The asymmetry $A_3$ which 
is independent of transverse polarization is found to be
\begin{eqnarray}
\displaystyle A_3^{CC}(\theta_0)&= 
 \displaystyle
       \; {\cal B}'_L 
 \frac{ \pi}{2}&\!\left[  (g_A-Pg_V)\left\{\sbar ({\rm Im} a_2+{\rm Im} a_5)+
2 {\rm Im} a_6   \right\} \right. \nonumber \\
& \;\;  & \left.\! \!+ (g_V-Pg_A)
 \left\{\sbar ({\rm Im} v_2+{\rm Im} v_5) +2{\rm Im} v_6 \right\}             \right] ,
    \end{eqnarray}
where
\begin{equation}
{\cal B}'_L =
\frac{{\cal B}_L (1-P_L\overline{P}_L)(\sbar - 1)\cos^2\theta_0}
        { (g_V^2+g_A^2-2Pg_Vg_A)\sigma_0^L},
\end{equation}
and 
\begin{eqnarray}
& \displaystyle \sigma_0^L=4 \pi {\cal B}_L(1-P_L\overline{P}_L)
  \left[ \left\{ \frac{\sbar^2 + 1}{(\sbar - 1)^2}
                   \ln \left( \frac{1 + \cmin }
                                   {1 - \cmin }
                            \right)
                 - \cmin
          \right\}\right]. \;
\end{eqnarray}
In accordance with the CPT theorem $A_3^{CC}$ is sensitive only to
imaginary parts of the couplings $v_2,\, a_2, \, v_5,\, a_5,\, v_6$
and $a_6$.

For the case of CV we have chosen two types of forward-backward
asymmetries, which happen to involve only 6 out of the 12 form-factors.
We define the following asymmetries:
\begin{eqnarray}
& \displaystyle A_1^{CV}(\thmin)=\frac{1}{\sigma_0^T}\sum_{i=0}^1 (-1)^n
\left(\int_{-\cos \theta_0}^{0} d \cos\theta-
\int^{\cos \theta_0}_0 d \cos\theta\right)
\int_{n \pi}^{(n+1)\pi} d\phi \,
{d \sigma \over d \Omega}\, , & \\
& \displaystyle A_2^{CV}(\thmin)=\frac{1}{\sigma_0^T}\sum_{i=0}^1 (-1)^n
\left(\int_{-\cos \theta_0}^{0} d \cos\theta-
\int^{\cos \theta_0}_0 d \cos\theta\right)
\int_{(n-1/2) \pi}^{(n+1/2)\pi} d\phi \,
{d \sigma \over d \Omega}\, &
\end{eqnarray}
which may be evaluated to read:
\begin{eqnarray}
& \displaystyle A_1^{CV}(\theta_0)= 
\displaystyle {{\cal B}_T\over 12 (g_V^2+g_A^2) \sigma_0^T}\cdot & \nonumber \\
& \displaystyle
\left((P_T-\overline P_{T}) g_A (
48 {\rm Re} s_5 + {\rm Re} s_6 (\sqrt{\sbar}-1)^2 \sbar (1+\sin^2\theta_0+
\sin\theta_0)) +  \right. &  \\
& \left. \displaystyle (P_T+\overline P_{T}) g_V (
48 {\rm Im} p_5 + {\rm Im} p_6 (\sqrt{\sbar}-1)^2 \sbar (1+\sin^2\theta_0+
\sin\theta_0))\right) \sbar (\sin\theta_0-1)+ & \nonumber \\
& \displaystyle  
{4{\cal B}_T\over (g_V^2+g_A^2) \sigma_0^T}(
(P_T-\overline P_{T}) g_V {\rm Im}s_1 
+(P_T+\overline P_{T}) g_A {\rm Re}p_1) \sqrt{\sbar} 
(\sin\theta_0-1)
& \nonumber 
\end{eqnarray}
and
\begin{eqnarray}
& \displaystyle A_2^{CV}(\theta_0)= 
\displaystyle {{\cal B}_T\over 12 (g_V^2+g_A^2) \sigma_0^T}\cdot & \nonumber \\
& \displaystyle
\left((P_T-\overline P_{T}) g_V (
48 {\rm Im} s_5 + {\rm Im} s_6 (\sqrt{\sbar}-1)^2 \sbar (1+\sin^2\theta_0+
\sin\theta_0)) +  \right. &  \\
& \left. \displaystyle (P_T+\overline P_{T}) g_A (
48 {\rm Re} p_5 + {\rm Re} p_6 (\sqrt{\sbar}-1)^2 \sbar (1+\sin^2\theta_0+
\sin\theta_0))\right) \sbar (\sin\theta_0-1)+ & \nonumber \\
& \displaystyle  
{4{\cal B}_T\over (g_V^2+g_A^2) \sigma_0^T}(
(P_T-\overline P_{T}) g_A {\rm Re}s_1 
+(P_T+\overline P_{T}) g_V {\rm Im}p_1) \sqrt{\sbar} 
(\sin\theta_0-1).
& \nonumber
\end{eqnarray}
The asymmetry $A_1^{CV}$ is sensitive to the real parts of the
couplings $s_5,\, s_6$ and $p_1$ and the imaginary parts of the
couplings $p_5,\, p_6$ and $s_1$, while the asymmetry
$A_2^{CV}$ is sensitive to the imaginary parts of the
couplings $s_5,\, s_6$ and $p_1$ and the real parts of the
couplings $p_5,\, p_6$ and $s_1$.  
Indeed, 
these asymmetries bring in the transverse $e^+$
and $e^-$ polarizations linearly in accordance with the
general considerations of Sec.~\ref{general}.

In the following section, we employ these asymmetries to provide
an estimate of sensitivities attainable at the linear collider
with realistic degrees of polarization and integrated luminosity.

\subsection{Sensitivities}
In this section, we shall present a brief discussion on the sensitivity that
can be obtained on the couplings at a linear collider with realistic
polarization and luminosity.  This is meant merely for the purposes of
illustration, and we shall provide a simplified discussion, assuming 
only one coupling nonzero at a time. We assume realistic values of
polarization and discuss the cases of same-sign and opposite-sign
polarizations separately. 
This has the advantage of isolating, e.g., the scalar and pseudo-scalar
couplings separately, etc..  Such a mode of operation is likely at the future
linear collider.
We shall separately present the sensitivities obtained for the
CV and CC cases in the following.
\EPSFIGURE[t]{Fig1.eps.new,width=13cm}
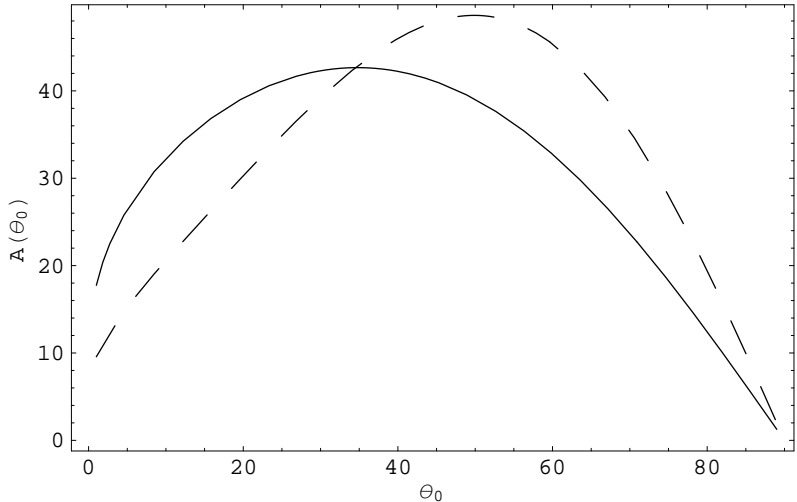
{Value of asymmetry $A_1^{CV}(\theta_0) (\times
10^3)$ obtained with Im$s_1=1$ (solid) and $A_2^{CV}(\theta_0)$ with
Im$s_6=1$ (dashed).}

\subsubsection{Chirality violating couplings}
In order to obtain the sensitivities of the CV couplings,
we make use of the asymmetries $A_i^{CV}$ ($i=1,2$).
We first consider the asymmetry $A_1^{CV}$ and assume only non-vanishing
imaginary parts.  We also assume that $P_T=0.8$, and $\overline P_T=-0.6$. 
For this simplified case, assuming Im$s_1=1$, and the
remaining Im$s_i$ to be vanishing, we can
compute the asymmetry, plotted as the solid profile in
Fig. 1.  The asymmetry $A_2^{CV}$ is the same with the
choice Im$s_5=1$, while the computed asymmetry with the choice Im$s_6=1$ is
given by the dashed profile in Fig. 1.

\EPSFIGURE[t]{Fig2.eps.new,width=13cm}
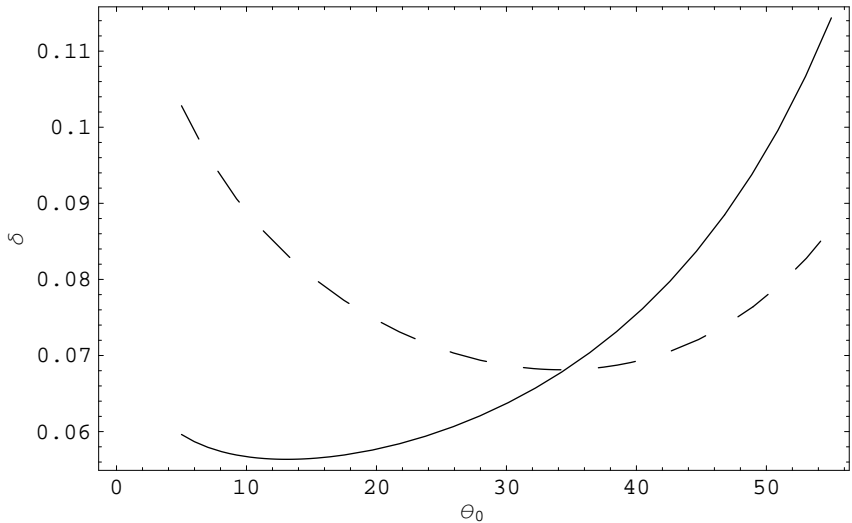
{Value of sensitivities from $A_1^{CV}$ on Im$s_1$ 
(solid) and from $A_2^{CV}$ on
Im$s_6 (\times 10^3)$ (dashed).}

We have here assumed
$\sqrt{s} = 500$ GeV, $\int L dt = 500$ fb$^{-1}$, and magnitudes of
electron and positron polarization to be 0.8 and 0.6 respectively.
We have calculated 90\% CL limits that can be obtained with a 
LC with the operating parameters given above.
Denoting this sensitivity by the symbol 
$\delta$ (\ie, the respective real or imaginary
part of the coupling), it is related to the value $A$ of 
a generic asymmetry for unit
value of the relevant coupling constant by:
\beq
\delta \equiv \frac{1.64}{|A|\sqrt{N_{SM}}},
\eeq
where $N_{SM}$ is the number of SM events\footnote{The coefficient
1.64 may be obtained from statistical tables for 
hypothesis testing with one estimator, see, 
e.g., Table 32.1 of ref.~\cite{PDG}.}.    
Our results are plotted in Fig. 2.
We may now optimize the sensitivity at the following angles, giving
us the following numbers: $|{\rm Im} s_{1,5}| \leq 5.6 \cdot 10^{-2},$
(optimum angle of $13^0$), $|{\rm Im} s_6| \leq 6.8 \cdot 10^{-5}$
(optimum angle of $34^0$).
These may be readily translated into sensitivities for the other couplings:
the sensitivity of Re$s_{1,5,6}$ is that of the corresponding imaginary
part multiplied by $g_V/g_A\simeq 0.08$ which yields:
$|{\rm Re} s_{1,5}| \leq 4.5 \cdot 10^{-3},$
(optimum angle of $13^0$), $|{\rm Re} s_6| \leq 5.4 \cdot 10^{-6}$
(optimum angle of $34^0$).
We keep in mind that these sensitivities are obtained by suitably
interchanging the asymmetries $A_1^{CV}\leftrightarrow A_2^{CV}$,
and switching the sign of the positron polarization.
Finally, we note that 
the sensitivities of the real and imaginary parts of the $p_{1,5,6}$
is identical to those of their scalar counterparts.

\subsubsection{Chirality conserving couplings}
We now come to 
the asymmetries and sensitivities in the CC case, which were already 
reported in ref.~\cite{ar2}.
We take up for illustration the case when only  ${\rm Re}~v_6$ 
is nonzero, since the results for other CP-violating combinations 
can be deduced from this case. 
We choose $P_T=0.8$ and $\overline{P}_T=0.6$, and vanishing longitudinal 
polarization for this case. 
Fig. 3 shows the asymmetries $A_i^{CC}$ as a 
function of the cut-off 
when the values of the anomalous couplings ${\rm Re}~v_6$ (for the case of 
$A_1^{CC}$) and Im~$v_6$ (for the case of $A_2^{CC}$ and $A_3^{CC}$) alone
are set to unity.
We have again calculated 90\% CL limits that can be obtained with a 
LC with $\sqrt{s} = 500$ GeV, $\int L dt = 500$ fb$^{-1}$, $P_T = 0.8$, and
$\overline{P}_T = 0.6$ making use of the asymmetries $A_i^{CC}$ ($i=1,2$).
For $A_3^{CC}$, we assume unpolarized beams.
The curves from $A_1^{CC}$ corresponding to setting only Re~$v_6$ nonzero, 
and from $A_2^{CC}$ and $A_3^{CC}$ corresponding to keeping only Im~$v_6$ 
nonzero 
are illustrated in Fig. 4.
That there is a stable plateau for a choice of
$\theta_0$ such that $10^{\circ} \lsim \theta_0 \lsim 40^{\circ}$; 
and we choose the optimal value of
$26^0$ (we note here that the angle is the same for
all cases considered, unlike in the CV case).  
The sensitivity corresponding to this for 
Re $v_6$ is $\sim 3.1\cdot 10^{-3}$.
The results for the other couplings may be
inferred in a straightforward manner from the explicit example above.
For the asymmetry $A_1^{CC}$, if we were to set $v_3(v_4)$ to unity,
with all the other couplings to zero, then the asymmetry would be
simply scaled up by a value $\sbar/2$, which for the case at hand is
$\simeq 14.8$.  The corresponding limiting value would
be suppressed by the reciprocal of this factor.
The results for the couplings ${\rm Re}~a_{2,5,6}$, compared to
what we have for the vector couplings 
would be scaled by a factor $g_V/g_A$ for the asymmetries
and by the reciprocal of this factor for the sensitivities.
The results coming out of the asymmetry $A_2^{CC}$ are such that the 
sensitivities of the imaginary parts of
$v$ and $a$ are interchanged {\it vis \`a vis} what
we have for the real parts coming out of $A_1^{CC}$.
The final set of results we have is for the form-factors that
may be analyzed via the asymmetry $A_3^{CC}$,  
which depends only on longitudinal polarizations. We treat the cases
of unpolarized beams and longitudinally polarized beams with $P_L = 0.8$, and
$\overline{P}_L = -0.6$ separately.  For the unpolarized case,
the results here for Im $v_6$
correspond to those coming from $A_2^{CC}$, with the asymmetry scaled
up now by a factor corresponding to $\pi/2$ and a further factor
$(P_T \overline{P}_T)^{-1}$ $(\simeq~2.1)$, which yields an overall factor
of $\sim 3.3$.  The corresponding sensitivity is smaller is by
the same factor.  Indeed, the results we now obtain for ${\rm Im}v_{3,4}$
are related to those obtained from $A_2^{CC}$ for $i=2,5$ by the same
factor.
For the case with longitudinal polarization, the sensitivities for the 
relevant ${\rm Im}v_i$ are enhanced by almost an order of magnitude, whereas the
sensitivities for ${\rm Im}a_i$ are improved marginally. 

\EPSFIGURE[t]{Fig3.ps.new,width=13cm}
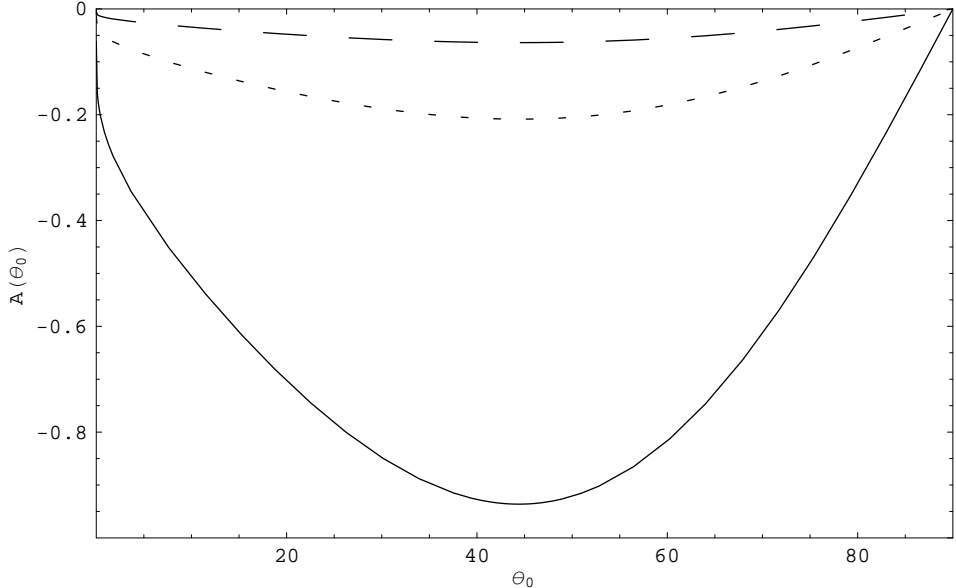
{The asymmetries $A_1^{CC}(\theta_0)$ (solid line), 
$A_2^{CC}(\theta_0)$ (dashed line) and 
$A_3^{CC}(\theta_0)$ (dotted line), defined in the text, plotted as functions
of the cut-off $\theta_0$ for a value of Re $v_6 ={\rm Im~}v_6 = 1$.}
\EPSFIGURE[t]{Fig4.ps.new,width=13cm}
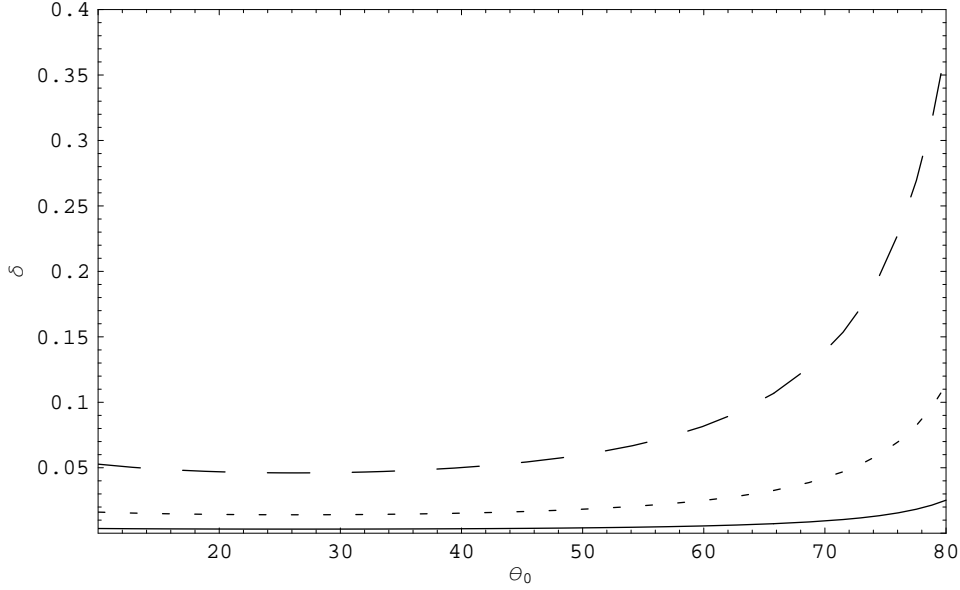
{The 90\% C.L. limit on Re~$v_6$ 
from the asymmetry $A_1^{CC}$ (solid line), 
and on Im~$v_6$ from $A_2^{CC}$ (dashed line) and  
$A_3^{CC}$ (dotted line),
plotted as functions of the cut-off $\theta_0$.}

In summary, from the asymmetry $A_1^{CC}$, we get
$|{\rm Re}v_{3,4}|\leq 2.1 \times 10^{-4},\, |{\rm Re} v_6|\leq 3.1\times
10^{-3}$, and
$|{\rm Re}a_{3,4}|\leq 3.1 \times 10^{-3},\, |{\rm Re} a_6|\leq 4.6\times
10^{-2}$, while the asymmetry $A_2^{CC}$ yields the sensitivities
$|{\rm Im}a_{3,4}|\leq 2.1 \times 10^{-4},\, |{\rm Im} a_6|\leq 3.1\times
10^{-3}$, and
$|{\rm Im}v_{3,4}|\leq 3.1 \times 10^{-3},\, |{\rm Im} v_6|\leq 4.6\times
10^{-2}$.
The asymmetry $A_3^{CC}$ which can be defined for unpolarized and
longitudinally polarized beams, yields the following sensitivities
for unpolarized (longitudinally polarized) cases:
${\rm Im}v_{2,5} \leq 9.3 \times 10^{-4}(5.6 \times 10^{-5})$,
${\rm Im}v_6 \leq 1.4\times 10^{-2}(8.4\times 10^{-4})$ and
${\rm Im}a_{2,5} \leq 6.4 \times 10^{-5}(5.2 \times 10^{-5})$,
${\rm Im}a_6 \leq 9.6\times 10^{-4}(7.9\times 10^{-4})$. 

It must be noted that the various couplings which are dimensionless
arise from terms that are suppressed by different powers of $m_Z^2$.
In particular, these must be viewed as model independent estimates,
which could be used to constrain specific models.

\section{Discussion and conclusions}\label{discconc}

It is now pertinent to ask what sort of models might lead to
such CC and CV form-factors.  In the context of the former, it
was already shown in~\cite{ar2} that anomalous triple-gauge boson
vertices\footnote{Here we do not give an exhaustive bibliography
for the work done on this source of CP violation and refer instead
to the references listed in a recent work~\cite{Ots}.} 
generate precisely the kind of correlations as $a_6,\, v_6$
provided we suitably identify the parameters.  In the context of
CV form-factors, Higgs models of the type considered in ref.~\cite{dw}
could give rise to couplings involving the $\epsilon$ symbol we have
considered here. 
In ref.~\cite{CSRR} CP even trilinear gauge boson
vertices at one-loop in the SM and minimal supersymmetric model (MSSM)
are computed, while the implications for colliders is considered in,
e.g., ref.~\cite{GLR}. An earlier work discussing the implications of
several different theoretical scenarios on the cross sections and
angular distributions in $e^+e^- \to Z \gamma$ is 
\cite{Mery:1987et}.


In conclusion,
we have considered in all generality the role of chirality conserving as well
as chirality violating couplings due to physics beyond the standard model.
The results due to the latter are entirely new and complement the former.
We started out by elucidating the role of longitudinal as well as
transverse polarization in phenomena such as these.  By relating the
number of independent helicity amplitudes to the possible number of
CP violating (and conserving) amplitudes, we narrowed down the a linearly
independent set of form-factors that could contribute to the differential
cross-section.  We have pointed out that for the case of chirality violation
it is sufficient to consider only scalar type terms, as the tensor like
terms are redundant, and is proof is provided in Appendix \ref{appa}.
We have constructed suitable asymmetries and have discussed their properties.
These asymmetries have been employed to provide estimates for the level
at which the new physics contributions may be constrained at the linear
collider with realistic polarization and integrated luminosity.  Our work
provides a comprehensive set of expressions which can be used at future
realistic detector environments and can be used to constrain and
calibrate realistic detector signals.

\setcounter{section}{0}
\renewcommand{\thesection}{\Alph{section}}
\section{Redundancy of tensor interactions}\label{appa}

Consider the massless Dirac equation,
\begin{equation}
\slp_1 u(p_1) = 0 ; \overline{v}(p_2) \slp_2=0.
\end{equation}
and the definition for the spin operator, 
\begin{equation}
\vec{\Sigma}= \gamma_5 \gamma_0 \vec{\gamma}.
\end{equation}
The following are then readily obtained:
\begin{eqnarray}
& \gamma_5 u(p_1) = (\vec{\Sigma} \cdot \hat{p}) u(p_1) & \\
& \overline{v}(p_2) \gamma_5 = \overline{v}(p_2) (\vec{\Sigma}\cdot \hat{p}) &
\end{eqnarray}
where $\hat{p}\equiv \vec{p_1}/|\vec{p_1}| = - \vec{p_2}/|\vec{p_2}|$.

For a tensor matrix (i.e. anti-symmetrized product of two 
gamma matrices) T, then
\begin{eqnarray}
& \overline{v}(p_2) \gamma_5 T u(p_1) =\overline{v}(p_2) (\vec{\Sigma} \cdot 
\hat{p}) T u(p_1), & \\
& \overline{v}(p_2) T \gamma_5  u(p_1) =\overline{v}(p_2)T (\vec{\Sigma} \cdot 
\hat{p})   u(p_1). & 
\end{eqnarray}
Adding these two equations, and using the fact that $\gamma_5$ commutes 
with T, we now obtain
\begin{equation}
2 \overline{v}(p_2) \gamma_5 T u(p_1) = \overline{v}(p_2) \{\vec{\Sigma}.
\hat{p}, T\} u(p_1).
\end{equation}
We note here that the anti-commutator on the right hand side 
is an anti-commutator of two commutators of $\gamma$ matrices. Well known
identities involving such anti-commutators and commutators can
be utilized to those that these reduce to combinations involving   
either 4 $\gamma$'s (i.e. the product of the $\epsilon$-symbol
with $\gamma_5$) or no $\gamma$'s. Thus the left hand side
reduces to a combination of a
pseudo-scalar and scalar. The exercise can be carried out 
by multiplying, at an earlier stage, by $\gamma_5$ yielding
\begin{equation}
2 \overline{v}(p_2) T u(p_1) = \overline{v}(p_2) \gamma_5 \{\vec{\Sigma}
\cdot \hat{p}, T\} u(p_1),
\end{equation}
and the same conclusion follows. 

\section{CP properties of anomalous couplings}\label{appb}
In order to establish the CP properties for the couplings 
in a transparent manner,
it is useful to consider the effective Lagrangian for the CC and the CV cases.
These are chosen to be Hermitian for the case that the 
couplings are purely real.
These read as follows:  
\begin{eqnarray}
& \displaystyle {\cal L}_I^{CC}={e^2\over 4 \sin\theta_W \cos\theta_W m_Z^4} \cdot & \nonumber \\
& \left\{-i\left[\partial_\mu \bar{\psi}(V_1+A_1 \gamma_5) \gamma_\beta \partial_\alpha \psi -
\partial_\alpha \bar{\psi}(V_1+A_1 \gamma_5) \gamma_\beta \partial_\mu \psi\right]-  \right. &\nonumber \\ 
& \displaystyle \left. i\left[ \bar{\psi}(V_2+A_2 \gamma_5) \gamma_\alpha \partial_\beta \partial_\mu \psi- 
\partial_\beta \partial_\mu \bar{\psi}(V_2+A_2 \gamma_5) \gamma_\alpha \psi \right] -  \right. & \nonumber \\
& \displaystyle \left. i\left[\partial_\beta \bar{\psi}(V_3+A_3 \gamma_5) \gamma_\alpha \partial_\mu \psi -
\partial_\mu \bar{\psi}(V_3+A_3 \gamma_5) \gamma_\alpha \partial_\beta \psi\right]+  \right. &\nonumber \\ 
& \displaystyle \left. \left[ \bar{\psi}(V_4+A_4 \gamma_5) \gamma_\alpha \partial_\beta \partial_\mu \psi+ 
\partial_\beta \partial_\mu \bar{\psi}(V_4+A_4 \gamma_5) \gamma_\alpha \psi \right] +  \right. & \nonumber \\
& \displaystyle \left. \left[\partial_\beta \bar{\psi}(V_5+A_5 \gamma_5) \gamma_\alpha \partial_\mu \psi +
\partial_\mu \bar{\psi}(V_5+A_5 \gamma_5) \gamma_\alpha \partial_\beta \psi\right]-  \right. &\nonumber \\ 
& \displaystyle \left. m_Z^2 \bar{\psi} (V_6+A_6\gamma_5) \gamma_\alpha g_{\beta\mu} \psi \right\}F^{\mu\alpha} Z^\beta, &
\end{eqnarray}
and
\begin{eqnarray}
& \displaystyle {\cal L}_I^{CV}={e^2\over 4 \sin\theta_W \cos\theta_W} \cdot & \nonumber \\
& \displaystyle \left\{ {1\over  m_Z^3} \bar{\psi} (S_1+i P_1 \gamma_5) \psi F_{\mu\alpha} \partial^\mu Z^\alpha + \right. & \nonumber \\
& \displaystyle \left. {1\over m_Z^5} \left[\partial_\mu \bar{\psi}(S_2+i P_2 \gamma_5) \partial_\alpha\partial_\beta \psi +
 \partial_\alpha\partial_\beta \bar{\psi}(S_2+i P_2 \gamma_5)\partial_\mu \psi\right]F^{\mu\alpha}Z^\beta + \right. & \nonumber \\
& \displaystyle \left. {i\over m_Z^5}\left[\partial_\mu \bar{\psi}(S_3+i P_3 \gamma_5) \partial_\alpha\partial_\beta \psi -
 \partial_\alpha\partial_\beta \bar{\psi}(S_3+i P_3 \gamma_5)\partial_\mu \psi\right]F^{\mu\alpha}Z^\beta + \right. & \nonumber \\
& \displaystyle \left. {1\over 2m_Z^3} \bar\psi (S_4+i P_4\gamma_5) \psi \epsilon_{\alpha\beta\rho\sigma}
		F^{\rho\alpha}\partial^\sigma Z^\beta - \right. & \nonumber \\
& \displaystyle \left. {i\over 2 m_Z^3} \left[\bar\psi (S_5+i P_5\gamma_5) \partial^\sigma \psi -\partial^\sigma \bar\psi(S_5+i P_5\gamma_5)\psi
					\right]\epsilon_{\alpha\beta\rho\sigma}
		F^{\rho\alpha} Z^\beta + \right. & \nonumber \\
& \displaystyle \left. {i\over m_Z^7}\left[\partial^\tau \partial_\mu \bar\psi(S_6+i P_6 \gamma_5) \partial^\sigma \partial_\alpha \psi+
\partial^\tau \partial_\alpha \bar\psi(S_6+i P_6 \gamma_5) \partial^\sigma \partial_\mu \psi\right] \epsilon_{\beta\rho\sigma\tau}
		\partial^\mu F^{\rho\alpha}Z^\beta \right\} & 
\end{eqnarray}
The CP properties of various terms in the above Lagrangians may be
determined using the standard CP transformation properties of the
electron, photon and $Z$ fields. The result is that in ${\cal
L}_I^{CC}$,                                    
the terms corresponding to $V_{1,2,3}$ and $A_{1,2,3}$ are CP even
and the rest are CP odd.
In ${\cal L}_I^{CV}$, the terms corresponding to $S_{1,2,5,6}\, {\rm and}\, 
P_{3,4}$ are CP even, while $S_{3,4} \, {\rm and}\,  P_{1,2,5,6}$ are CP
odd.

Once the CP properties are established from the effective Lagrangians, 
it is no longer necessary
to restrict the $V_i, \,A_i, \,S_i, \,P_i, \, i=1,...6$ 
to be real, and we allow them to be complex.  
In accordance with the CPT theorem, the correlations arising from 
these will be CPT-even (odd) for the real (imaginary) parts of the 
form-factors.

Introducing plane-wave solutions for the $e^+$, $e^-$, $Z$ and $\gamma$
corresponding to the momenta given in eq.~(\ref{process}),
and promoting the coupling constants 
$V_i,\, A_i,\,i=1,...,6$
to complex form-factors dependent on kinematic invariants $s$, $t$ and
$u$,
we find that to reproduce the vertex factor of eq.~(\ref{anomcc}),  
we need to make the following replacements:
$V_1\to i v_1,\, 
V_2\to i(v_2-v_5), V_3\to i(v_3-v_4),\, V_4\to (v_2+v_5),\,
V_5\to (v_3+v_4), V_6=v_6$, and 
$A_1\to i a_1,\, 
A_2\to i(a_2-a_5), A_3\to i(a_3-a_4),\, A_4\to (a_2+a_5),\,
A_5\to (a_3+a_4), A_6=a_6$. 
Analogously, for the CV case, 
again with
$S_i,\, P_i,\,i=1,...,6,$ taken to be form-factors, 
we need to make the following replacements to 
reproduce
eq. (\ref{anomcv}):
$S_1\to s_1,\, S_2\to (s_2-s_3)/2, S_3\to -i(s_2+s_3)/2,\, 
S_4\to s_4,\,
S_5\to -is_5, S_6\to -is_6$ and
$P_1\to p_1,\, P_2\to (p_2-p_3)/2, P_3\to -i(p_2+p_3)/2,\, 
P_4\to p_4,\,
P_5\to -ip_5, P_6\to -ip_6$. 
The CP properties of the momentum-space form-factors,
which are stated explicitly in the text and tabulated in
Tables 1 and 2 may be easily deduced from
the respective CP properties of the couplings in the Lagrangian.

\acknowledgments{BA thanks the Council for Scientific and 
Industrial Research for support during the course of these
investigations under scheme number 03(0994)/04/EMR-II.}

\end{document}